\documentclass[5p]{elsarticle}

\usepackage{hyperref} 

\usepackage{tabularx}

\usepackage{pbox}

\journal{Physics Letters A}

\begin{document}

\begin{frontmatter}

\title{Perspectives of Using Oscillators for Computing and Signal Processing} 

\author{Gy\"orgy Csaba, } \address{Faculty for Information Technology and Bionics P\'azm\'any P\'eter Catholic University}

\author{Wolfgang Porod} \address{Center for Nano Science and Technology, University of Notre Dame}

\begin{abstract} 
It is an intriguing concept to use oscillators as fundamental building blocks of electronic computers. The idea is not new, but is currently subject to intense research as a part of  the quest for 'beyond Moore' electronic devices.  In this paper we give an engineering-minded survey of oscillator-based computing architectures, with the goal of understanding their promise and limitations for next-generation computing. We will mostly discuss non-Boolean, neurally-inspired computing concepts and put the emphasis on hardware and  on circuits where the oscillators are realized from emerging, nanoscale building blocks. Despite all the promise that oscillatory computing holds, existing literature gives very few clear-cut arguments about the possible benefits  of using oscillators in place of other analog nonlinear circuit elements. In this survey we will argue for  finding the rationale of using oscillatory building blocks and call for benchmarking studies that compare oscillatory computing circuits to level-based (analog) implementations.
\end{abstract}


\end{frontmatter}


\section{Introduction}

The goal of this paper is to give a survey of oscillatory computing architectures (OCAs) and help the reader to understand the value and utility of this concept in engineering applications, and  in the design of beyond Moore computing and signal processing devices.

The vast majority of signal processing / computing devices uses a level-based representation of signals - in most cases, the  levels are discrete or binary voltage values. In these models of computation the dynamic properties of the signals play a secondary role -- the information that is carried by the timing of the signals or by their exact waveform, is largely neglected. Wasting the information carried by the timing of the signal inevitably wastes energy - this suggests that any purely level-based computing scheme is suboptimal in terms of power consumption, which is arguably the most important figure of merit in todays computing devices.

 Information is physical and in any hardware realization, the signal  must be represented by a physical quantity (such as current, voltage,  or an other emerging variable).  One could almost certainly do better and use the carriers of information more efficiently. One way of accomplishing this is to use the phase and /or frequency of oscillatory signals to carry information, besides the signal amplitude. This is one of the key ideas behind OCAs.

The other key characteristics of an OCA is that the signals have to be processed by the physical interaction of nonlinear oscillators. There are many emerging computing concepts that use the interaction of nonlinear elements for computation, \cite{ref:cnn}  - but, as it turns out, it is not at all easy to find physically realizable low-power, robust, nonlinear elements that could serve as building blocks in such computers. Oscillators are good candidates for a practical nonlinear devices  - one reason for that is that  oscillatory systems are ubiquitous in the physical world, so one may hope to find an attractive (low power, compact, fast, etc) device realizations. Most physical oscillators show a suitable nonlinearity in their phase or frequency response and may act as the nonlinear, active circuit element at the heart of the computing architecture.

\begin{figure}[!ht] 
\centering 
\includegraphics[width=3.1in]{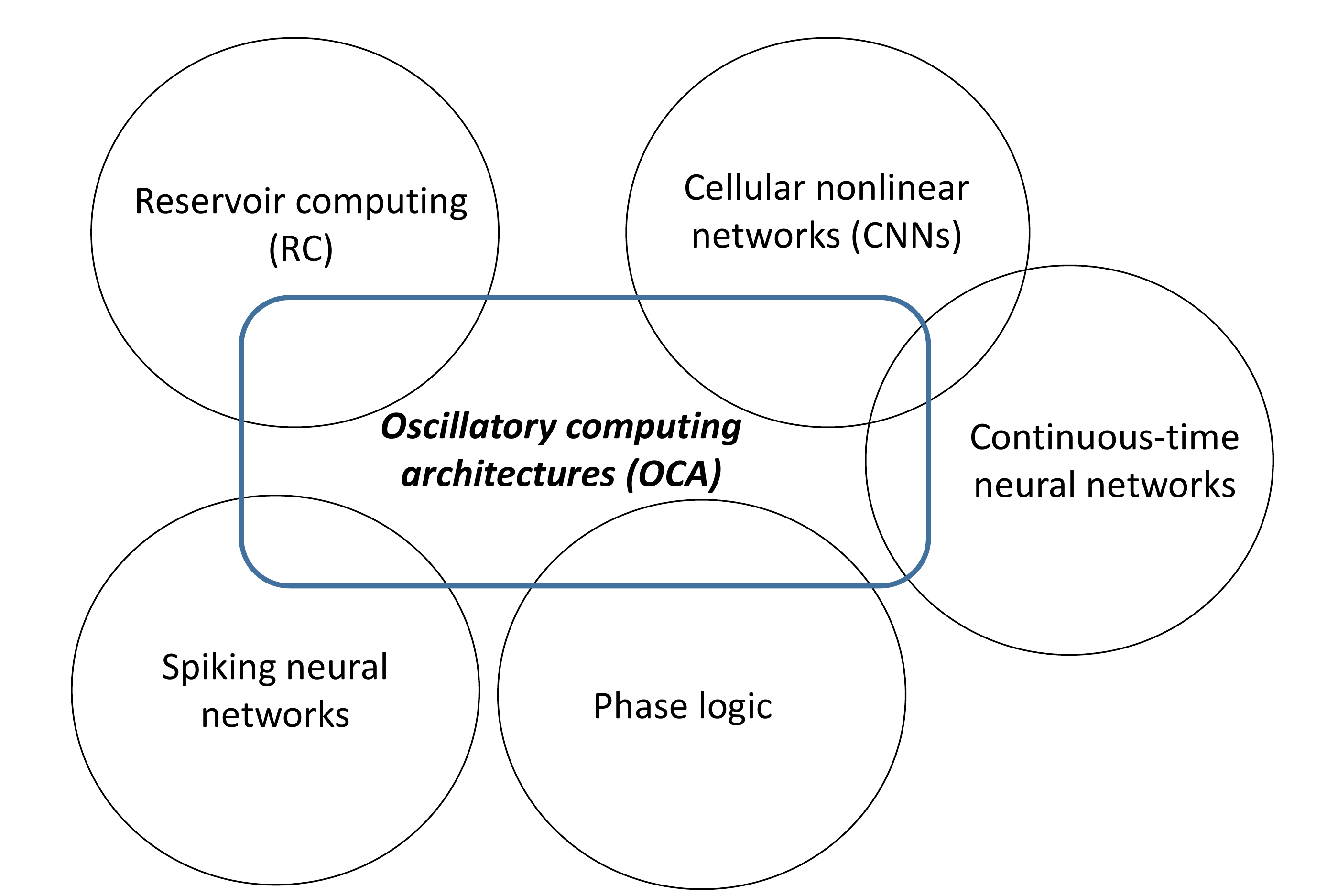} 
\caption{Oscillatory computing architectures are closely related or overlap with other analog computing concepts - some of those relationships are illustrated above.} 
\label{fig:fig_overview} 
\end{figure}

Figure \ref{fig:fig_overview} gives a schematic high-level overview of analog, dynamic computing concepts, that are relevant for OCAs.  The field of OCAs overlaps with many nonconventional computing paradigms and many analog, non-conventional computing model can be implemented using oscillatory building blocks.

Von Neumann's oscillatory computer \cite{ref:vonneumann}, is an early and still very relevant example of OCAs - this scheme uses phases of oscillatory signals to realize Boolean, digital computation. Von Neumann's oscillatory computer serves as a perfect example how one can redesign a level-based computing scheme (in this case a standard Boolean computer) to phase / frequency  / based realization and also illustrates most  of the challenges inherent to this approach. For this reason  we will  review this concept in Section \ref{sec:vonneumann}.

Many different kinds of oscillators and oscillator coupling schemes can be implemented in transistor-based (CMOS-based) circuitry, but such circuits will inherit many disadvantages of 'standard' CMOS circuitry - in particular, power figures will likely be \emph{not} be orders of magnitude better than in CMOS digital devices. Using emerging devices and possibly, non-electrical state variables \cite{ref:statevariables} would have the potential to be yield to truly revolutionary devices.  We will argue that the physical realization of the oscillators is pivotal for their success in real-life application and devote Section \ref{sec:physics} to physics of emerging nano-oscillator devices.


Biologically inspired OCAs(oscillator-based neuromorphic computing) represent a major area within the field \cite{ref:rojas} and we cover this area in Sec. \ref{sec:bio}. But with the recent triumphs of artificial neural networks (ANNs, such as deep learning \cite{ref:deng_deeplearning} \cite{ref:deeplearning} \cite{ref:cmos_deep} and related ideas) a more engineering-minded approach is emerging to oscillator-based devices, that is, to use them as hardware accelerators for repetitive, power hungry operations in ANNs.  Prime example here are image-preprocessing operations in an image processing pipeline \cite {ref:upside1} associative memory units, reservoirs of an echo state network \cite{ref:grollier_new} or units handling computationally hard problems \cite{ref:suman_coloring}. Our review will devote Sec. \ref{sec:special} and Sec. \ref{sec:np} to these new application areas.

The reader will see that oscillatory computing devices is a vast field - the 100+ reference we cite in this review represents only a small fraction of the literature and there are books devoted to the topic \cite{ref:osc_nn}. Still,  it seems that a fundamental 'why' question remains unanswered, namely, why one would use oscillatory components instead of other nonlinear circuit building blocks? It is very much possible to realize analog, neuromorphic, non-Boolean computing devices from non-oscillatory nonlinear elements  -- much researched Cellular Nonlinear Networks (CNNs) \cite{ref:cnn} do excatly that and memristor-based, non-oscilatory neural  neuromorphic architectures are a hot topic nowadays \cite{ref:memristor_neural}. We mentioned above that oscillators may be attractive due to the vast number of possible implementations - but this alone is a rather hand-waving argument as oscillators have obvious disadvantages too. For example they must run continusly, dissipating power all the time, and it can be energetically costly to power them up / down - there must be strong benefits to offset such disadvantages. We believe that OCAs indeed may have fundamental benefits and in the conclusion of this paper we offer clues for what those killer benefits could be.

Before diving in the details of oscillatory computing we have to make an important distinction between OCAs and  spiking neural networks / spike-based signal processors. Spiking (neural) networks are also an emerging field and perhaps the first successful engineering application related to neurally-inspired circuit architectures. The boundary between OCAs and spiking networks is somewhat diffuse and the terminology is often imprecise. Spiking neural networks employ oscillatory signal representation (they count spikes, use their relative phase and frequency to do processing), but they do \emph{not} rely on oscillator interactions: most architectures count / integrate pulse sequences using digital circuits or analog integrators.  Their promise comes from the fact that a single spike can carry extremely low energy (in the order of $10^{-16}$  J) and it is acceptable to lose some fraction of the spikes \cite{ref:upside1} \cite{ref:spike_survey}  \cite{ref:spike_overview_1} \cite{ref:spike_overview_2}  \cite{ref:spike_overview_3}  \cite{ref:upside2} \cite{ref:izhikevich_improc}\cite{ref:izhikevich_spike} \cite{ref:spike_querlioz}  \cite{ref:essderc} \cite{ref:STDP_cruz_albrecht} to noise.  Spiking neural networks may use oscillators for generating the signals but the nonlinear interaction, pattern formation of these oscillators is not required for their operation. Spiking devices are not subject of this review paper.

\section{Phase logic and von Neumann's oscillatory computer} \label{sec:vonneumann}

Von Neumann's idea \cite{ref:vonneumann} \cite{ref:neumann_patent}  is based on interconnected, subharmonic injection locked oscillators (SHILOs).  While we cannot be sure of von Neumann's motivation, it is worthwhile to note that digital computes in the early fifties reached then-breathtaking several MHz speed already in the early fifties. So it seemed a natural idea to look at a logic circuit not as  switch between zeros and ones, but rather as an electrical oscillator switching between different phases of oscillation. Goto \cite{ref:goto}  and others furthered this concept (the basic oscillator elements are called parametron in their works \cite{ref:parametron1} ) and fully functional Boolean computers were realized using these building blocks \cite{ref:musasino}.

\begin{figure}[!ht] 
\centering 
\includegraphics[width=3.1in]{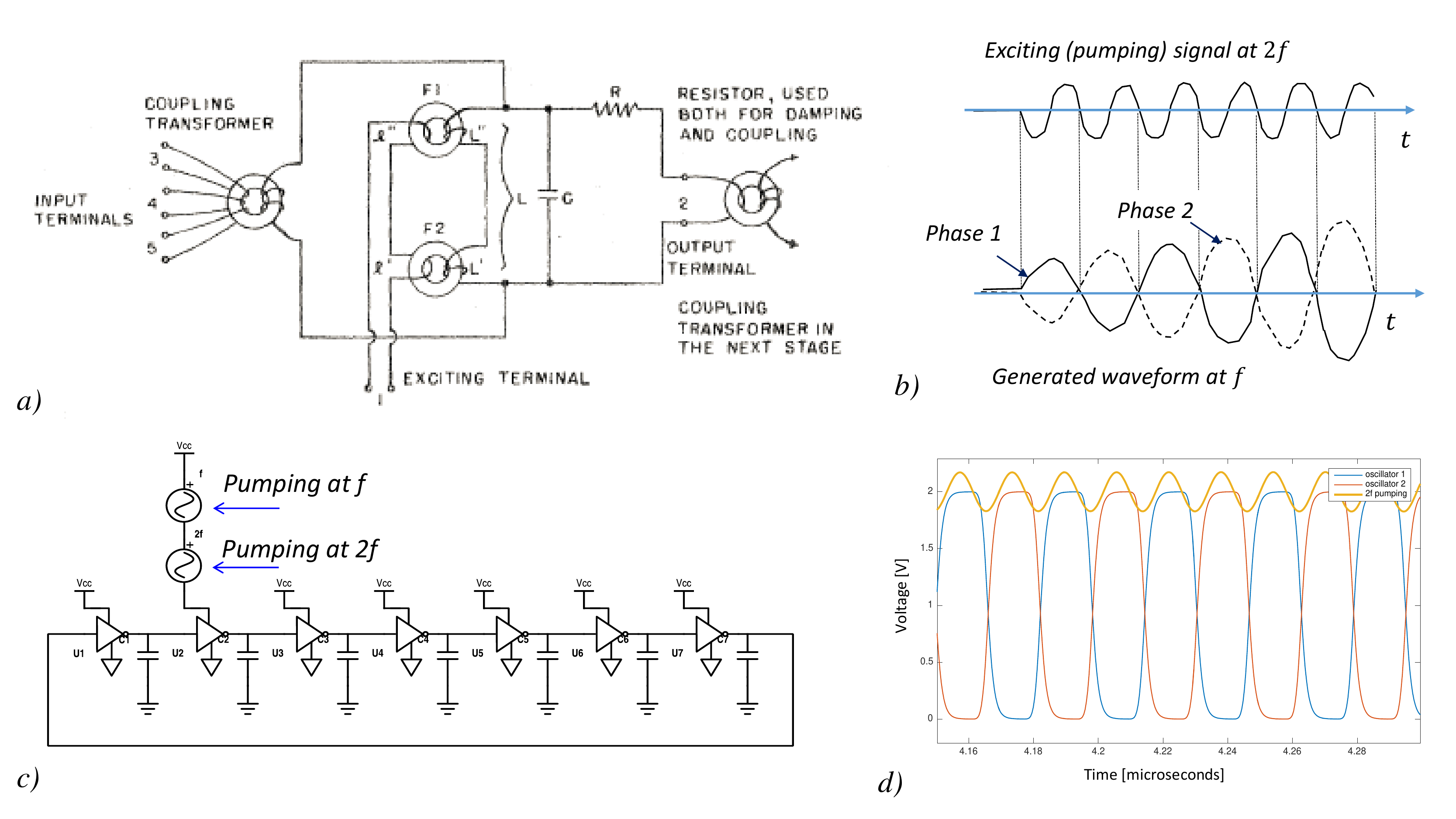} 
\caption{$a)$ Circuit schematics of a magnetic ($LC$ oscillator-based) parametron, taken from \cite{ref:goto} $b)$ illustrates how a $2f_0$ excitation gives rise to $f_0$ frequency oscillations. $c)$ shoes a more integration-friendly variation of the idea with ring oscillators, along with the simulated waveforms in $d)$} 
\label{fig:parametron} 
\end{figure}

Figure \ref{fig:parametron}$a-b$ is a circuit schematics of a logic gate from a parametron-based computer. This is one particular implementation, where the building blocks are inductively coupled nonlinear $LC$ oscillators and the oscillator nonlinearity comes from the nonlinear hysterezis of the inductive cores. If the $LC$  oscillators have a resonance frequency of $f=f_0$ and are excited by an external $f_\textrm{pump}=2f_0$ signal. This parametric pumping supplies the oscillators with energy (they need no DC power supply), and  the $f=f_0$ frequency oscillator signals become phase-locked to $2f_0$ pumping signal. One may see in Figure \ref{fig:parametron}b that there are two distinct phases in which a $f_0$ frequency signal may be synchronized to $2f_0$ signal.  This observation is the key for using this system for binary logic  - the two distinct phases with respect to the pumping signal, represent the binary '0' and '1'.

Multiple, interconnected oscillators (which all run at $f_0$ frequency in one or the other possible phase)  will pull toward each other's phase. If a particular oscillator receives multiple inputs with different phases, then it will follow the phase of the \emph{majority} of input oscillators. Such majority gates serve as universal logic gates in the computing architecture  - they can straightforwardly realize NAND / NOR gates and inverters and satisfy the five tenets of Boolean computation \cite{ref:nikonov_benchmarking}.

Clearly, $LC$ oscillators have serious practical limitations: inductors remain a major challenge in any planar and miniaturized technology. The idea of phase-based logic however, is resurrected in many proposals. Ring oscillators may be a microelectronic-friendly implementation \cite{ref:rowch} \cite{ref:phlogon_book} - Fig. ref{fig:parametron}$b-c$ illustrates this possibility.  Using nanomechanical oscillators   \cite{ref:nanomech3} \cite{ref:parametron1} one could make very compact and extremely low-power stand-alone logic gates - so the idea of parametron is very much alive. One can construct all phase logic elements from single electron devices \cite{ref:kiehl_phaselogic}. It is argued \cite{ref:rowch}  that phase-based representation are more noise tolerant than level-based signal representation, which can be extremely important in low-power, low-voltage circuitry.

Oscillator phases could represent analog variables, but parametric pumping reduces the of stationary phase states to two, turning the SHILOs into a digital system.  This makes the scheme realization friendly - it is not necessary to use high-quality oscillators with stable frequency and the phase characteristics, the external stimulus stabilizes phases / frequencies. In the original scheme the analog computing power of oscillators is not harnessed - albeit one may extend the idea to non-Boolean, analog computation \cite{ref:montecarlo}, as illustrated in Fig.  \ref{fig:parametron}b).

Parametron-type devices offered equivalent functionality to level-based logic gates, but they never became mainstream due to the lack of scalable, fast, low-power, on-chip oscillators. In these years, Moore's law  \cite{ref:moore}, in its original 'scale everything' form is ending, due to power constraints. Oscillator-based logic (phase logic) may have a chance for a comeback, if it can reach significantly better power figures, either with electrical \cite{ref:rowch} or with alternate state variables \cite{ref:statevariables}.


\section{Hardware realizations for oscillatory circuits} 

\label{sec:physics}

Just like as a digital computer is built from billions of transistors, an envisioned OCA will contain millions or billions of interconnected oscillators. The demands for the elementary oscillators are high and the success of OCAs will eventually depend on whether one can find oscillatory building blocks that are (1) compact (2) low lower (3) high-frequency (4) low noise (5) can be efficiently interconnected to each other (6) and easily interfaces with electronic circuitry.

Satisfying all of the above requirements is a tall order - and, depending on the chosen  architecture, some of the requirements may just be unimportant.  Table \ref{tab:oscils} below gives an overview of possible physical oscillators, and we also give quick estimation of some important parameters. Some of these oscillators are electrical \cite{ref:oscillator_book}, some use another state variable in between their electrical contacts, others may do every function with non-electrical state variables  \cite{ref:statevariables}.

\begin{table*}[h] \footnotesize \centering \resizebox{\textwidth}{!}{ \begin{tabular}{ | c | c | c | c | c | c |} 

\hline Oscillator name & state variable	& frequency [Hz] & power dissipation  /  cycle  [J]  & possible coupling mechanism & remarks, references, for further probing \\ 

\hline Ring oscillator & Electric         	& up to 20 GHz & $10^{-15}$									  & electrical only				 	  & \cite{ref:lowpowerVCO2} \\ 
\hline 

Relaxation oscillator based on phase-transitions						    & electrical        & up to 10 GHz 				& $10^{-17}$ 					 & electrical l 		&  \cite{ref:Shukla2014IEDM} \\ 
 \hline 

LC oscillator	  								& Electrical & up to 100 GHz   			& 								  & electrical only				 	  &  \cite{ref:inductors_excellent} \\ 
\hline

Superconducting oscillator				    & Electric, magnetic       & several 10 GHz 				& $10^{-17}$ 									 & electrical, inductive, capacitive 		& \cite{ref:josephsonsync1} \cite{ref:josephsonsync2} \\ 
 \hline 

 Mechanical (NEMS) oscillator / RBO  			& Mechanical 		& up to 20 GHz &  $10^{-14}$  											  & electrical or mechanical 		 & \cite{ref:nanomech_1} \\ 
\hline 

 Spin torque oscillator (STO) 					& magnetic 		& upwards 50 GHz 		     & $10^{-15}$ 											  & electric, magnetic or spin wave &  \cite{ref:grollier_spintorque} \cite{ref:kaka_pufall} \\
 \hline 

 Chemical   										& electrochemistry 	& $10^2$ 					     & no data & no data &  \cite{ref:chemical} \\ 
 \hline \pbox{3cm}{Magnetic Anisotropy controlled  parametric}  & magnetic 		& up to 20 GHz 				&  		no data									 & electrical  						&  \cite{ref:krivorotov} \\ 
 \hline 

Spin-Hall  oscillator						    & magnetic        & up to 20 GHz 				&  $10^{-16}$ 											 & electrical or mechanical 		& \cite{ref:akerman9sto} \\ 
 \hline

\hline
 \end{tabular} } 
 \label{tab:oscils} 
\caption{Possible building blocks of an oscillatory computing architecture. Ring oscillators can be viewed as good baseline for any emerging device concept. }
 \end{table*}

\subsection{Figures of merits for various oscillators}

An ideal oscillator  in terms of energy would be an energy-recycling oscillator, i.e. where during each oscillation cycle energy is converted (largely reversibly) between two forms, instead of being dissipated. An $LC$ oscillator converts between electrostatic and magnetic energies, a mechanical (NEMS) oscillator converts between kinetic and potential energy \cite{ref:nanomech_1} \cite{ref:nanomech_2} \cite{ref:nanomech3} \cite{ref:nanomech_4} \cite{ref:mechanical_new}. Such oscillator with a quality factor $Q$ dissipates only  $1/Q$th of its stored energy. Unfortunately, both $LC$ and NEMS oscillators, which have this property \cite{ref:highQcmos}, fare quite badly in the other figures of merits \cite{ref:mems_efficiency}. The $LC$ oscillators require  inductors, which are very hard to integrate on chip, they are bulky and resistive. So on-chip  $LC$ oscillators have low $Q$ factors \cite{ref:inductors_excellent} and little can be gained by the energy recycling. Mechanical oscillators (MEMS / NEMS) can have very high $Q$s - in their case the transduction efficiency is the main problem: no matter how efficient, high $Q$ oscillators they are, interconversion between electric and mechanical signals can be rarely done better than with a few-percent efficiency. Many NEMS oscillators that were proposed for computation \cite{ref:nanomech_4} occupy a large chip area - albeit there are a lot of new developments toward practical nanomechanical oscillators \cite{ref:rbo_new}.

In terms of sheer numbers, superconducting devices may come closest to the a perfect oscillators: they consume ultra low energies, capable of high frequencies,  they do not necessarily occupy extreme chip areas, and to some degree they are energy-recycling, just as the $LC$ circuits \cite{ref:supercond00} \cite{ref:josephsonsync1} \cite{ref:josephsonsync2} \cite{ref:supercond_oscillator}. Their obvious disadvantage is the required cooling apparatus, they are challenging to integrate with input / output circuitry, memory units, and it is not all clear, how easy it would be to construct a highly interconnected network using them. For these reasons they do not belong to the mainstream of research now.

There are many variants of electrical oscillators one may use. Ring oscillators are one of the most compact transistor-based oscillators  - their frequency and power consumption can vary in a wide range and in subthreshold mode they can compete with low-power nanoscale oscillators \cite{ref:ROneuron} \cite{ref:ringoscillator_old}. Interconnections to each other and interfaces to electronic circuitry are straightforward.

Electrical oscillators may be built from emerging devices, such as phase change materials -- these are often referred to as memristive oscillators  \cite{ref:corinto_memristor}. These operate as relaxation oscillators, the oscillation frequency is set by the $RC$ time constant of the oscillator \cite{ref:Shukla2014IEDM}. The switching element itself is a very simple structure and highly scalable.

Chemical oscillators \cite{ref:chemical} provide an interesting link to biological systems - but perhaps their use in integrated computing systems are limited.

A large class of oscillators relies on oscillatory motion of magnetic materials (spin excitations). Spin torque oscillators and spin-hall oscillators use spin-polarized current to excite spin-precession in a ferromagnetic layer, and due to some form of magnetoresistance (giant magnetoresistance (GMR) or via magnetic tunnel junctions (MTJs) the current flowing through the oscillator is modulated by the magnetization oscillations. The magnetic thin film is capable of high-frequency, high-$Q$ oscillations, but the interconversion between the electric and magnetic degrees of freedom  is not particularly high and for these reasons, magnetic oscillators are not particularly low-power. Their high oscillation frequency makes them outstanding candidates for computing and they are one of the most popular physical realizations for computing \cite{ref:grollier_new} \cite{ref:sto_1}  \cite{ref:grollier_spintorque} .

It is also possible to build parametric oscillators from magnetic layers - using voltage-controlled anisotropy is a particularly efficient way of doing that \cite{ref:krivorotov} - these lend themselves naturally to the realization of a von-Neumann type oscillatory computer \cite{ref:krivorotov} \cite{ref:parametric_magnet}.

\subsection{Power considerations}
 
A good baseline for comparing the power dissipation of various oscillators is to calculate the power figures for ultra-low power ring oscillators, which are used in, for example RFID transponders \cite{ref:lowpowerVCO} \cite{ref:lowpowerVCO2}. The voltage-controlled oscillator described in \cite{ref:lowpowerVCO2} consumes 24 nW at 5.24 MHz, that is $E_{\mathrm diss}=4.7\times 10^{-15}$ J per oscillation cycle. With vanadium oxide relaxation oscillator one can go to an order of magnitude better: \cite{ref:parihar} projects $0.5 \mu$ W at 1.6 GHz,  giving $E_{\mathrm diss} \approx 10^{-16}$ J per cycle/
 
 Spin torque oscillators are current-driven devices and  typically run at sub-millamperes current and at GHz frequency - their voltage (and the dissipated power) depends on how low the resistance of the stack can be mode. Assuming $V_{\mathrm STO}=0.1 $ V, $i{\mathrm STO}=0.1$ mA and $f=10 $ GHz, the energy consumed per cycle is $E_{\mathrm diss} \approx 4.7\times 10^{-15}$ J per cycle, not very far from the ring oscillator figure - and in case if STOs one has to deal with the losses to / from conversion to electrical signals. Other magnetic resonators (spin-hall effect based, voltage-controlled) may fare better. 
 
Superconducting devices stand out with a $E{\mathrm diss} \approx 10^{-17}$ J energy per spike. 
 
The energy of thermal fluctuations at room temperature is $\mathrm{k}T=26$ meV $=4.1430\times 10 ^{-21}$ J. The energy involved in each oscillation cycle should be at least a few times this value to avoid the oscillator signal getting completely lost in noise. The above presented oscillators are still several orders of magnitude away from this value - so there is certainly room for more efficient hardware.

 \subsection{The role of oscillator noise}

It is in general true that low-energy, nanoscale oscillators have noisy waveforms, with relatively poor frequency and phase stability \cite{ref:RO_phasenoise} \cite{ref:maffezoni_models} \cite{ref:sumandatta}. In non energy-recycling oscillators one has to keep the energy of the signal as low as possible, in order to minimize dissipation. For example, the energy contained in a small-size STO is in the order of hundreds of room temperature k$T$s - significantly higher than the characteristic energy of thermal fluctuations, but not as high as to completely avoid thermal noise to influence oscillator waveforms \footnote{In magnetic oscillators, the magnetic energy stored in the device may get fairly close to $\mathrm{k}T$ - unfortunately, this does not translate to low power dissipation, due to the poor efficiency of magneto-electric interconversion}. In relaxation oscillators or ring oscillators, low values of the charge-storing $C$ capacitor will lead to such regime.

In general, the phase of nanoscale oscillators will jiggle, while the frequency remains more stable. Computing schemes relying on the average frequency or average phase of oscillators in a network are more experimentally viable than schemes that require long-term phase coherence \cite{ref:nikonov01}. Noise itself  may be useful for the computation - it may strengthen the interaction of weakly coupled oscillators \cite{ref:hanggi} and may eliminate metastable, local energy minima.

An extreme case of noisy oscillators are stochastic oscillators \cite{ref:grollier_stochastic}. In case of magnetic oscillators stochastic behavior is achieved by reducing the volume of the magnetic free layer and as a result reducing the energy of the oscillator to the $\Delta E \approx 10$ k$T$ range. In this case, the oscillator spontaneously jumps between states, requiring no energy for the switching (albeit energy still will be required to interconnect or read out the oscillators). Some computational schemes may exploit such fairly random oscillations.

\subsection{Physical coupling of oscillators}

In order to turn them into computing networks, oscillators must be interconnected, which can be done electrically or using an emerging state variable, such as the electrical or magnetic degree of freedom. In a high-interconnection architecture (such as a most neuromorphic schemes) the number of interconnections greatly outweighs the number of oscillators. So one may argue that the figure of merits for oscillators as given in Table \ref{tab:oscils} are not at all that relevant, and 'good' oscillators are the ones that can be coupled by compact, low-power, high-fanout interconnections.  Even in standard digital circuits, interconnections often account for most of the circuit complexity and moving data between far-lying points of the circuit accounts for most of the power consumption. It is not hard to see, that in highly interconnected analog (oscillatory) circuitry interconnections will be the realization bottleneck.

\begin{figure}[!ht] 
\centering 
\includegraphics[width=3.1in]{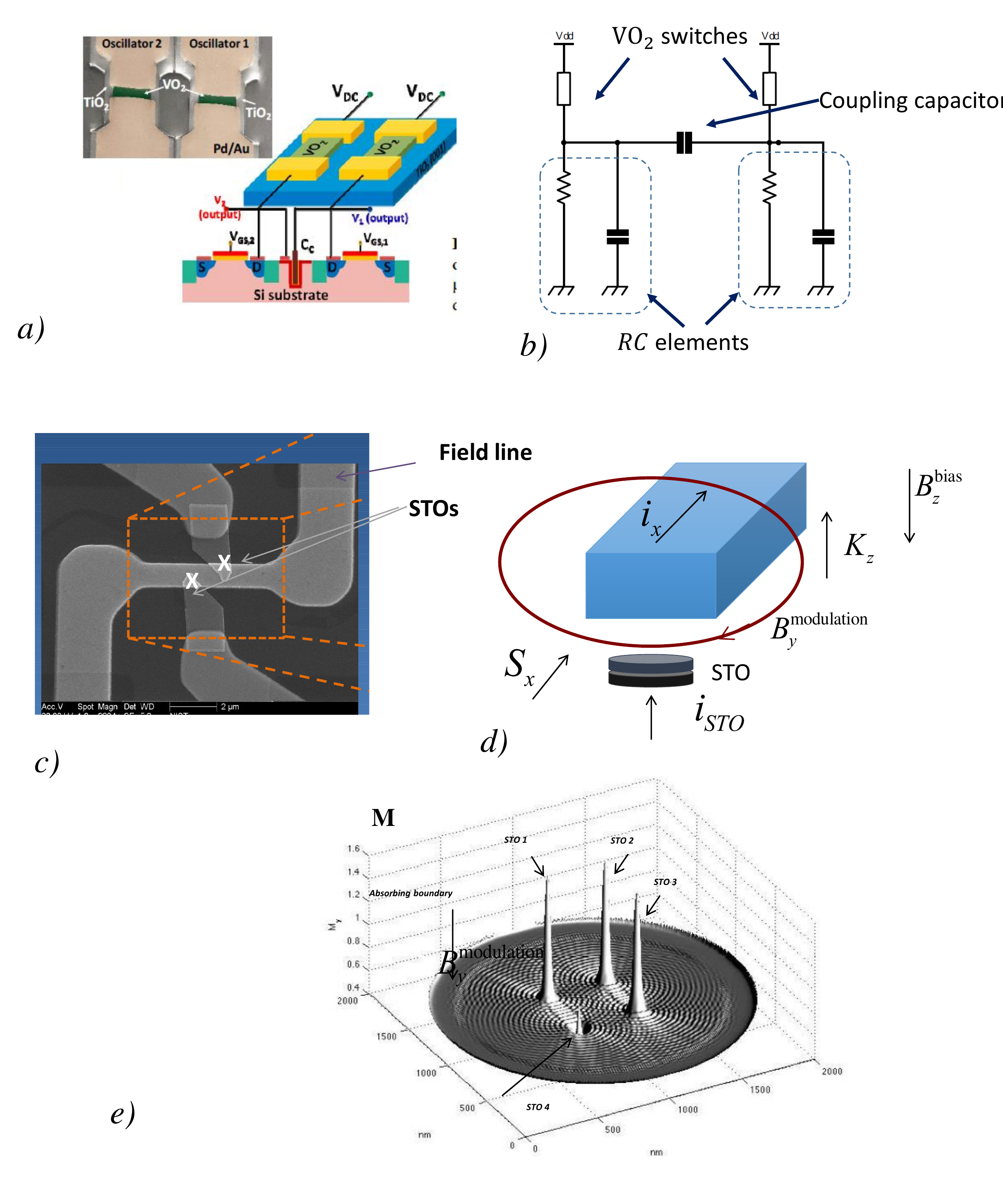} 
\caption{Three examples of oscillator interactions. $a)$-$b$ Electrical relaxation oscillators may be interconnected by capacitors, taken from \cite{ref:Shukla2014IEDM}. For STOs one may use electrical connections ($c)$-$d)$ \cite{ref:csaba_pufall_ieee}, which requires signal transduction and external amplifiers. Direct spin wave coupling avoids this problem - $d)$ shows four interconnected STOs with the spin wave pattern around them.} 
\label{fig:stocoupling} 
\end{figure}

Electrical interconnection is a straightforward  choice for coupling electrical oscillators  \cite{ref:montecarlo} \cite{ref:sumandatta}. The strength of electrical interconnections may be tunable or hard wired and often one needs both positive of negative coupling coefficients (i.e. ones that push or pull the phases against / toward each other). Figure \ref{fig:stocoupling}$a)-b)$ gives two examples for electrical oscillator interconnection, for $VO_2$-based relaxation oscillators.

Electrical connections are also the usually the easiest, most flexible option for oscillators operating on different state variables. One example of a relatively simple coupling scheme y is shown in Fig. \ref{fig:stocoupling}c)-d). In this case, high-frequency spin-torque oscillators (STOs) are coupled to a field line - which is simple a wire, providing magnetic field for the STOs. The field line current may be controlled by either an external current source (providing a control signal to the wire) or a signal that is provided by the STOs itself and brings them into mutual interaction. 

The structure of Fig. \ref{fig:stocoupling}c), in fact already illustrates many of the challenges associated with electrical interconnections of emerging oscillators. STO signals are relatively weak, so one must pick up and amplify their signal by a relatively heavy electrical circuitry (this is not shown in Fig. \ref{fig:stocoupling}) The big advantage of STOs is that they operate at several GHz frequencies, creates many challenges in routing the signals, one has to design appropriate microwave circuitry for routing the signals to the STOs

Stronger electrical signals between the STOs make the feedback amplifiers redundant and allow passive interconnection schemes for mutual coupling, this was was demonstrated in \cite{ref:electrical_sto}.

Since the interconversion between non-electric and electric degrees of freedom will inevitable result in some overhead, it seems highly desirable to use emerging state variables for interconnections as well. In the case of NEMS oscillators, this may be done by mechanical (acoustic) coupling \cite{ref:mems_acoustic}. In case of spin oscillators, dipole (magnetic field) or spin-wave coupling may work \cite{ref:kaka_pufall} (see Fig. \ref{fig:stocoupling}$e)$ - so far, up to  nine oscillators were brought to interaction this way\cite{ref:akerman9sto}. It is also possible to sync spin oscillators via the current distributions they create, without the need for additional circuitry. Physical coupling of emerging oscillators is probably the one that enables to fully utilize the potential of emerging oscillators, but the geometry and the damping of acoustic / spin waves strongly limits the realizable coupling schemes.

\section{Computing by oscillator dynamics, interconnection schemes} 
\label{sec:math}

Oscillatory computing schemes are realized by interconnecting a number of oscillators and exploit their collective dynamic behavior  for doing computation. The input of the network could be the initial phase of the oscillators or it could be encoded in the  frequency of (variable frequency) oscillators, or the inputs can be external modulating signals. In most cases the output is a stationery  phase,  frequency or amplitude pattern. This pattern may be the result of a Boolean computation or the solution of an image processing, optimization, pattern matching problem.

The functionality of the network is defined by the oscillator interconnections and / or the parameters of the oscillators or by external stimuli.  Mathematically, oscillator interaction are interpreted in terms of oscillator synchronization, i.e. the emergence of phase / frequency patterns in the oscillator cluster \cite{ref:mems_hoppensteadt} \cite{ref:hoppensteadt_pll} \cite{ref:pikovsky}. Theoretically, the simplest model to describe the formation of these states is is the celebrated Kuramoto model \cite{ref:kuramoto} \cite{ref:kuramoto_big}. A comprehensive overview of different neuron and coupling models is given in \cite{ref:izhikevich_weakly}. In the engineering / physics community 'oscillatory ground state'  is often used to describe the convergence of the dynamics system toward a useful, stationary pattern, and this pattern most often minimizes an energy function.

In practice, one often needs time-demanding numerical simulations or numerical approximations to determine phase dynamics \cite{ref:levitan_phase_model} - analytical models are not always useful to describe irregularly connected networks or oscillators that describe highly nonlinear physical systems.

\begin{figure}[!ht] 
\centering 
\includegraphics[width=3.1in]{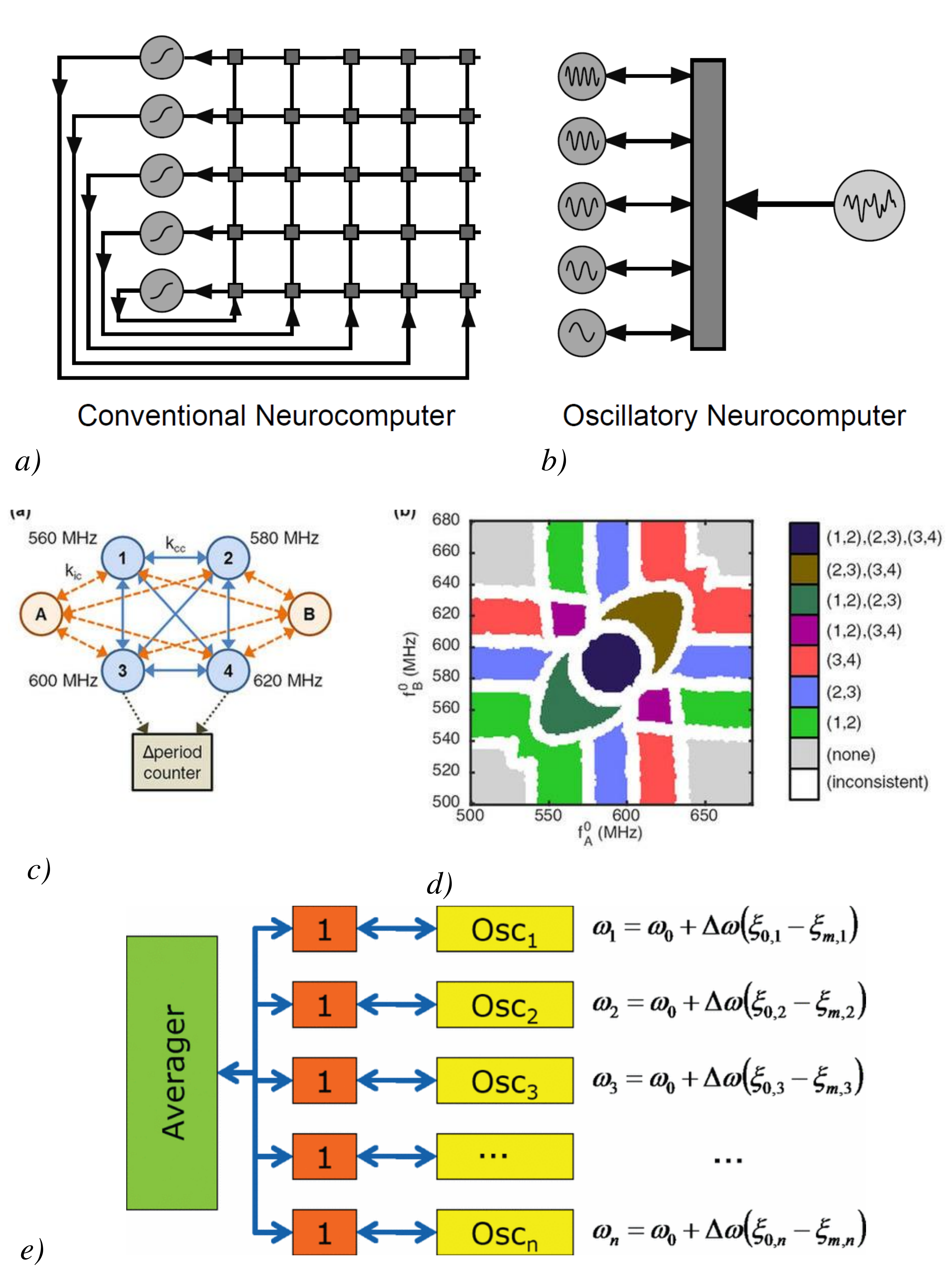} 
\caption{A few possible schemes for interconnecting oscillators into a useful computing network.  $a)$ is the classical scheme for an all-to-all interconnected, Hopfield-like net, oscillators enable to realize this network functionality with much fewer interactions, using exploiting frequency-division mulitplexing $b)$, both figures are taken from \cite{ref:izhikevich}. Using sinusoidal signals to control the oscillators, one can train the network of $c)$ to exhibit distinct synchronization patterns ($d)$, both figures are taken from\cite{ref:querlioz}.). The panel of $d)$ (from \cite{ref:nikonov01}) sketches a simpler interconnections scheme, where frequency controlled oscillators interact via a common node, the output of the computatio being the degree of scynchronization betwee the oscillators.  } 
\label{fig:suppes} 
\end{figure}

In many cases, one looks at the emergence of a phase pattern, i.e. clusters of phase-correlated oscillators. An important point, which is often lost in the mathematical definitions of synchronization that it does \emph{not} necessarily mean perfectly in-phase or anti-phase running oscillators - any (phase) correlation between two oscillators could have computational value and such phase correlation may be possible possible to access via an output circuitry.  An illustration for quasi synchronization (correlated phase states) is given in Fig. \ref{fig:suppes}, \cite{ref:suppes} \cite{ref:querlioz}. This pattern was generated by a 3-oscillator all-to-all coupled oscillator array, which is controlled by externally applied sweeping oscillatory signals. In the time domain, the behavior of such system seems almost like noise - extracting the pairwise phase correlations and after a threshold criteria, one gets the pattern of  Fig. \ref{fig:suppes}.

On the implementation level one may follow two distinct routes to define the functionality of the network. It is possible to rely on physically defined, often random couplings oscillators - in these case the oscillator parameters (frequency) or externally injected signals may be used to engineer the function of the network and / or a fully programmable outer layer can define the network function. This latter route is followed, for example in reservoir computing \cite{ref:neural_reservoir} \cite{ref:grollier_new}. 

Alternatively, one may engineer the interconnections (the coupling scheme) of the network for a certain functionality. Most oscillator coupling schemes are derived from a particular artificial neural network (ANN). For example, the well-established Hopfield network \cite{ref:hopfield} \cite{ref:porod_hopfield} is the starting point of oscillatory associative memory concepts, such as \cite{ref:izhikevich} and \cite{ref:montecarlo}. Couplings in such networks can be computed from a Hebbian rule, in the simplest case. The energy function of a well-known, level-based Hopfield network can be rewritten for phase signals and one can construct the oscillatory version of the Hopfield net (see \cite{ref:izhikevich}). This is perhaps the most established route \cite{ref:bogdan}. 
No matter which route is followed, one may use learning schemes developed for neural networks to determine the parameters of the oscillatory neural network \cite{ref:suppes} \cite{ref:querlioz}.

Most neurally inspired networks are highly interconnected or all-to-all connected, which always raises realizability issues and which was already a major bottleneck in traditional neural network architectures. Oscillatory signals will yield to many parasitic couplings so one may speculate that they are well-suited to highly interconnected systems, if these parasitic couplings can be exploited for useful purposes.

\section{Biologically inspired network models and deep learning nets by oscillators} \label{sec:bio}

The central nervous system is believed to use time-dependent spiking signals to communicate and process information - the dynamic nature of information processing in the brain is what probably distinguishes it most from today's digital computers. There is a diffuse boundary between spiking circuits and oscillators, consequently, many oscillatory networks are inspired by neuromorphic analogies or brain models. Excellent examples of using biological models to develop computational architectures are given by \cite{ref:baldi} and \cite{ref:furber_temple}. A comprehensive overview of circuits dedicated for neural computation is given in \cite{ref:superimportant}  One may look at the problem the other way: it is a intriguing idea to use special-purpose hardware to simulate complex neural processes and perform for example  large-scale brain simulation \cite{ref:largescalebrain}, albeit we are not aware of using oscillatory computers in this way.

Any man-made hardware that attempts to imitate neural systems faces an interconnection bottleneck. Neurons have a fan-out on the order of $10^5$, i.e. this is the number of direct point-to-point interconnections emanating from a single neuron in the brain. Microelectronic technologies cannot even come close to this number. One way to create a highly interconnected network is to use frequency-division multiplexing (FDM) in artificial neural networks and use a single, high-bandwidth physical link to create a large number of channels between processing units (neurons) \cite{ref:yuminaka}, \cite{ref:fm_cnn} \cite{ref:fm_neural} \cite{ref:fm_neural2} \cite{ref:izhikevich_weakly} \cite{ref:heger_krischer_new} \cite{ref:izhikevich} \cite{ref:horvath}. The possibility of using FDM provides another hand-waving argument in favor of oscillatory signal representation. Of course, there are other ideas to create highly interconnected network, such as  to mimic the internet infrastructure on chip  \cite{ref:marculescu} - but since this a high-level solution, it is likely to be suboptimal to circuit or device-layer solutions.

The field of neural networks were viewed by many as  an academic backwater until the past few years when deep learning networks started to revolutionize machine intelligence \cite{ref:ann_book}, \cite{ref:nn_book2} \cite{ref:goodfellow_deepleraning} \cite{ref:deng_deeplearning}  \cite{ref:deeplearning} \cite{ref:dengdongdeep}. Deep learning nets are now viewed by many as a panacea and a one fits all type of solution, where only the performance of available hardware limits what can be done. True or not,  deep learning nets require immense computer power especially in the training phase. Deep learning was a killer application that made many-core graphics-core (GPU) based processing mainstream, ending the 50-year dominance of single (or few) core CPU systems.  Clearly, deep-learning would be a strong motivation for special purpose hardware and \cite{ref:cmos_deep} and oscillator based solution would be welcome - we are not aware of efforts to this end.

\section{Special-purpose analog computing with oscillators} 
\label{sec:special}

In many computing tasks, vast majority of  resources (energy, time, hardware) is spent on relative simple,  repetitive jobs.  This is especially true in areas such as image processing, where large number of convolutions, filtering, image processing steps have to be done on massive amount of input data (i.e. video streams). Co processors for such tasks could be promising testbeds for oscillator-based computers since they target a well-defined task, which could easily be compared with existing, number-crunching solutions to the same problem. Since the OCA is acting as a hardware accelerator or co-processor in these tasks (not as stand-alone computing unit), it is especially important to estimate a net system-level improvements that may come from the OCA.

One of the outstanding problems is efficient preprocessing of video streams, i.e. filtering / convolution operations at the input side of an image processing pipeline (IPP). The focus of a recent DARPA project \cite{ref:upside_wired} targeted exactly such task \cite{ref:nikonov01} was the demonstration of  complete IPP, with analog oscillatory computinf devices at its heart. The underlying idea is that   efficient euclidean distance calculation on analog data can be done by expoiting oscillator interaction \cite{ref:tadashi}. The effort included  circuit design and algorithm design components and also a nanodevice  work package, where nanoscale mechanical and magnetic oscillators were developed as hardware components for the image processing pipeline (IPP).

The mechanism of oscillator interaction and how this is used for analog computing is described in \cite{ref:nikonov01} in detail. A much simplified sketch of the architecture was sketched in Fig. \ref{fig:suppes}e. The oscillator frequencies should be controllable via their driving current or voltage (VCOs) and their interactions should be sufficiently strong to pull the entire oscillator array in a synchronized state if their natural frequencies are sufficiently close. In fact, the presence or absence of the synchonized state indicates how close the oscillator frequencies are to each other. If the oscillator signals are superposed, rectified and integrated then the value of this output will be higher for synchronized states (where oscillator signals add up coherently) than for unsynchronized states. The network gives a measure of how the oscillator frequencies (i.e. the oscillator driving current or voltages) are clustered. This can directly be used for analog Euclidean distance calculation \cite{ref:upside1} \cite{ref:upside2} \cite{ref:kaushikroy}. The circuit itself requires all oscillators to be connected via a single control node. Distance calculations on $n$-element long vector require $n$ coupled oscillators.

\begin{figure}[!ht] 
\centering 
\includegraphics[width=3.1in]{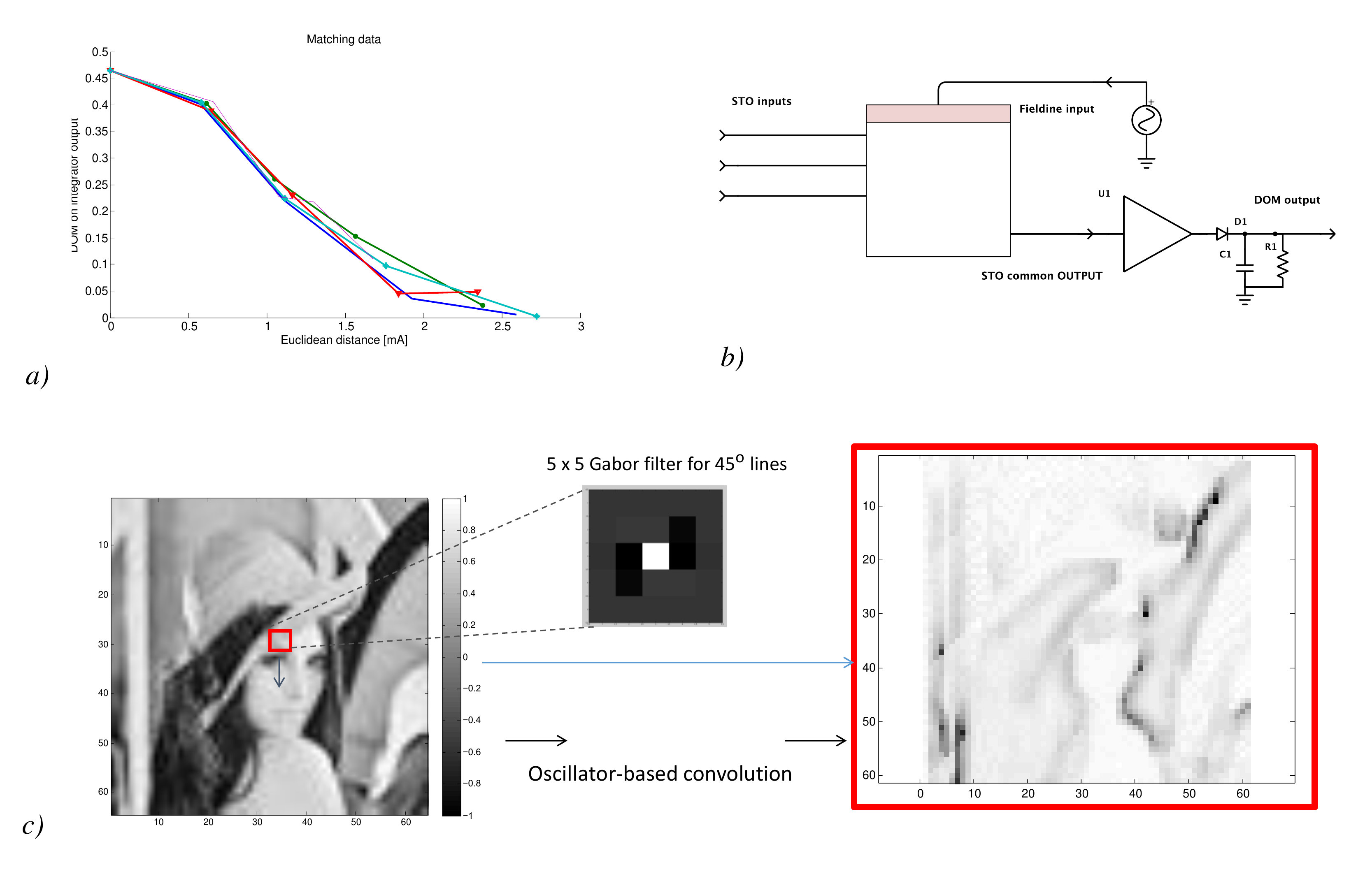} 
\caption{An image-processing pipeline based on oscillators. Euclidean distance calculation calculates dot products, which, in turn, can be used for Gabor filtering. $a)$ intergrator circuit $b)$ oscillator output vs. Euclidean distance $c)$ scanning through an image in patches.} 
\label{fig:nikonov} 
\end{figure}

One may wonder how this all comes together for a useful computing task - Fig. \ref{fig:nikonov} illustrates the ingredients for an IPP. The level of oscillator synchronization is measured is simply measured via a power detector (integrator) Fig. \ref{fig:nikonov}$a)$. The output of the integrator is to a good approximation linear with the Euclidean distance of the oscillator frequencies (Fig. \ref{fig:nikonov}$b)$ - the input  is measured in currents, since the used STOs are current-controlled oscillators). Using a  cluster of 25 couped STOs one can scan through an image in $5 \times 5$ patches - Fig. \ref{fig:nikonov}$c)$ demonstrates this scan for detecting 45-degree lines in the Lena image.

The oscillator unit can perform the analog distance calculation with potentially very high energy efficiency and due to the relative simplicity of the the coupling scheme, this circuit has recently become possible with nanoscale oscillators, such as electrically coupled STOs \cite{ref:csabasto} \cite{ref:pufall_new}. The relative simplicity of the computational task, however, turns out to be a disadvantage - as all the other image processing steps are done in the digital domain. The simple analog oscillator network  requires some sort of interface circuitry between the surrounding digital circuitry and and the  analog oscillator components. The high cost of A/D and D/A converters, current sources for driving the oscillators etc. and other input / output components  cannot be amortized. If the OCA targets a 'too simple' problem, then the benefits on the system level could be dwarfed by the high cost of interfaces to the conventional, digital components. For emerging oscillators this problem is usually more serious, as they often require more non-trivial interfaces to the digital world.

There are a number of other special-purpose computing tasks that had been demonstrated with oscillators, such as associative memory and pattern recognition tasks. Unlike the above described euclidean distance calculation, these problems can be scaled to operate on large, complex inputs \cite{ref:baldi} \cite{ref:syncspecial} and therefore the cost of input / output interfaces can be amortized. Nanoscale oscillators (STOs) were also demonstrated as building blocks \cite{ref:grollier_new} albeit it seems that the external circuitry and not the nanoscale oscillators do the heavy lifting in the computing process and potential system-level benefits are hard to estimate.

In standard neural implementations of the associative memory / pattern recognition networks the computational task is encoded in the interconnection weights of the neuraons and all-to-all interconnections are required. Such complex network of finely tuned connections does not lend itself to practical implementations. There are a few recent ideas for oscillatory neurons that are a lot more (nano)technology friendly. The work of \cite{ref:heger_krischer_new} uses the a frequency-domain multiplexing scheme to drastically reduce the number of inteconnections. Along a similar line of thought, the work of \cite{ref:grollier_new2} uses external oscillatory signals to control partial synchronization of oscillators and effectively control their interconnection strengths this way.

\section{Neural networks for hard problems} \label{sec:np}

As we argued above, a practically useful OCA should  target a sufficiently hard problem in order to mitigate the cost of I/O circuitry - ideally, a hard problem meaning that it is entirely intractable using digital circuitry or non-oscillatory networks.

Computationally hard problems (such as NP-hard problems) \cite{ref:karp21} \cite{ref:turinglimit} are usually discussed in the context of quantum information theory (quantum computation and quantum simulators). A quantum system in a fully entangled state can be described by exponentially growing number of superposition coefficients - i.e. the time evolution of $N$ coupled 2-state systems generally requires $2^N$ number of internal variables. A quantum computer or quantum simulator could store and operate simultaneously on exponentially large ($2^N$ sized) data - so a relatively small-sized hardware could, in principle, process vast-sized problems. Currently, large-scale industrial and academic efforts are ongoing to experimentally verify this claim \cite{ref:dwave}.

It is widely believed that no classical system, only quantum processors could do the feat of storing / processing information that is exponentially growing with the size of the system. Contrary to this common belief, it is quite possible that there is no fundamental difference between quantum and classical systems in this respect. This can be argued for  on three  grounds: (1) mathematically the relations between complexity classes (P, NP) are unknown  (2) the information content in collective states (excitations) of a classical system may grow exponentially with system size (3) eventually, classical and quantum systems will be subjected to the same type of noisy environments and may be limited by the same type of physical constraints. One should appreciate the significance and difficulty level of NP-hard problem (for an excellent insight, see \cite{ref:aaronson} ) and it is quite possible that  NP hard problems, in their general form, are beyond reach for both classical and quantum computers.

In the light of the above, however,  it makes sense to think about oscillator-based accelerators for NP-hard problems, i.e. to design OCAs that compete with \emph{realizable} quantum computers. The first attempt we are aware of addresses the graph coloring problem \cite{ref:graphcoloring_earliest}  \cite{ref:wu_graphcoloring}. It is not hard to see that the dynamics of in-phase and out of phase synchronizing oscillators map to a solution of a 2-color graph coloring problem, and even the oscillator interconnections directly correspond to the graph edges. A more general approach is shown in \cite{ref:suman_coloring}, together with an implementation using vanadium-oxide-based relaxation oscillators. In the approach of \cite{ref:suman_coloring} the problem is first reformulated to finding a circular ordering  of the nodes such that the same colored nodes appear together in the ordering - this preserves the hardness of the problem but turns it into a energy-minimization (optimization) task that oscillators can handle.

Another approach for using the collective states for exponentially hard probmens  is memcomputing \cite{ref:memcomputing_1} \cite{ref:memcomputing_2}  \cite{ref:memcomputing_3} \cite{ref:memcomputing_4}  \cite{ref:memcomputing_sciam} \cite{ref:memcomputing05} \cite{ref:memcomputing06}. This can be implemented in various physical systems, among them oscillators \cite{ref:memcomputing_1} - the number of collective excitations in a physical system grows quickly, enabling the solution of hard problems. The arguments of \cite{ref:memcomputing_1} are especially important, as they study the feasibility of the exponentially growing state space in the presence of noise. 

It must be noted that it is not exclusively OCAs that have been proposed to handle NP-hard problems, but other complex, non-oscillatory analog systems  \cite{ref:cnn_np_ercsey} \cite{ref:ercsey_sudoku} \cite{ref:ercsey_ksat} or memristors \cite{ref:memcomputing_3}. The prospect that complex analog dynamics systems may attack NP-hard problems could become the most important argument for their research and if this is proven true, all issues with analog / digital interfaces would become non-issue.  / 

\section{Conclusions and Outlook}

In this paper we gave a hardware-oriented review of the flourishing research field of oscillatory computing architectures (OCAs). Most works on the field are motivated by biological analogies (i.e. neuromorphic computing), or finding new applications for emerging hardware (such as a spin-oscillator-based computing systems). It would be highly useful and stimulating to substantiate fundamental rationale for the superiority of OCAs for certain problems types.

Finding a convincing and somewhat general argument for oscillatory computing system is likely not an easy task. A much older and still unsettled question is whether in general digital or analog solutions are superior for neuromorphic (bio-inspired) computational tasks \cite{ref:brocket_analog} - but there are many benchmarks and case studies out there. On the other hand, oscillator-based analog vs. level-based analog benchmarks are almost nonexistent.

We conclude with a few, somewhat hand waving arguments to corroborate the usefulness of OCAs. One of their benefits could be that they use of narrow-bandwidth device to device communication channels (i.e. oscillators running at a given frequency) allows efficient intra-circuit communication in the presence of noise\cite{ref:yuminaka}, \cite{ref:fm_neural2} \cite{ref:izhikevich}  and consequently, ultimately low-power operation could be accomplished \cite{ref:winograd}. A second possible  argument is that coupled oscillators are ubiquitous in the physical world, so one may find a 'perfect' device for a given computational task. A third, very encouraging fact is that interacting oscillators now appear in circuits proposed for the solution of NP-hard problems. It is very much possible that they will steal the show from quantum computing and yield to hardware that could handle  problems that seem intractable with today's resources.

\section{Acknowledgements}

The authors are grateful for George Bourianoff and Dmitri Nikonov for the motivation to join oscillator-based computing project, and for Matt Pufall and for Trond Ytterdal for excellent technical collaborations. We also acknowledge funding from the DARPA UPSIDE project and NSF NEB 2020.



\begin{thebibliography}{1}


\bibitem{ref:cnn} Roska, Tamas, and Leon O. Chua. "The CNN universal machine: an analogic array computer." IEEE Transactions on Circuits and Systems II: Analog and Digital Signal Processing 40, no. 3 (1993): 163-173.

\bibitem{ref:vonneumann} Wigington, R. L. A new concept in computing Proceedings of the IRE 47, no. 4 (1959): 516-523.

\bibitem{ref:statevariables} Kerry Bernstein, Ralph K. Cavin III, Wolfgang Porod, Alan Seabaugh, and Jeff Welser, Device and Architecture Outlook for Beyond-CMOS Switches, Proceedings of the IEEE 98 (12), 2169-2184 (2010).

\bibitem{ref:rojas} R. Rojas: Neural Networks: A Systematic Introduction Springer 1996

\bibitem{ref:deng_deeplearning} Deng, Li, and Dong Yu. "Deep learning: methods and applications." Foundations and Trends� in Signal Processing 7, no. 3?4 (2014): 197-387.

\bibitem{ref:deeplearning} LeCun Y, Bengio Y, Hinton G. Deep learning. Nature. 2015 May 28;521(7553):436-44.

\bibitem{ref:cmos_deep} Lu, Junjie, Steven Young, Itamar Arel, and Jeremy Holleman. "A 1 TOPS/W analog deep machine-learning engine with floating-gate storage in 0.13 �m CMOS." IEEE Journal of Solid-State Circuits 50, no. 1 (2015): 270-281.

\bibitem{ref:upside1} Hull, Zachary, Donald Chiarulli, Steven Levitan, Gyorgy Csaba, Wolfgang Porod, Matthew Pufall, William Rippard et al. "Computation with Coupled Oscillators in an Image Processing Pipeline." In CNNA 2016; 15th International Workshop on Cellular Nanoscale Networks and their Applications; Proceedings of, pp. 1-2. VDE, 2016.

\bibitem{ref:grollier_new} Jacob Torrejon,	Mathieu Riou,	Flavio Abreu Araujo,	Sumito Tsunegi,	Guru Khalsa,	Damien Querlioz,	Paolo Bortolotti,	Vincent Cros,	Kay Yakushiji,	Akio Fukushima,	Hitoshi Kubota, Shinji Yuasa,	Mark D. Stiles Julie Grollier Neuromorphic computing with nanoscale spintronic oscillators Nature 547, 428–431 (27 July 2017) doi:10.1038/nature23011

\bibitem{ref:suman_coloring} Parihar, Abhinav, Nikhil Shukla, Matthew Jerry, Suman Datta, and Arijit Raychowdhury. "Vertex coloring of graphs via phase dynamics of coupled oscillatory networks." Scientific Reports 7, no. 1 (2017): 911.

\bibitem{ref:osc_nn} M. G. Kuzmina, E. A. Manykin, E. S. Grichuk: Oscillatory Neural Networks  De Gruyter (November 15, 2013)

\bibitem{ref:memristor_neural} Pershin, Yuriy V., and Massimiliano Di Ventra. "Experimental demonstration of associative memory with memristive neural networks." Neural Networks 23, no. 7 (2010): 881-886.

\bibitem{ref:spike_survey} Paugam-Moisy, H�lene. Spiking neuron networks a survey. No. EPFL-REPORT-83371. IDIAP, 2006.

\bibitem{ref:spike_overview_1} Maass, Wolfgang. "Networks of spiking neurons: the third generation of neural network models." Neural networks 10, no. 9 (1997): 1659-1671.

\bibitem{ref:spike_overview_2} Paugam-Moisy, H�lene, and Sander Bohte. "Computing with spiking neuron networks." In Handbook of natural computing, pp. 335-376. Springer Berlin Heidelberg, 2012.

\bibitem{ref:spike_overview_3} Grning, Andr, and Sander M. Bohte. "Spiking Neural Networks: Principles and Challenges." In ESANN. 2014.

\bibitem{ref:upside2} Pufall, Matthew R., William H. Rippard, György Csaba, Dmitri E. Nikonov, George I. Bourianoff, and Wolfgang Porod. "Physical implementation of coherently coupled oscillator networks." IEEE Journal on Exploratory Solid-State Computational Devices and Circuits 1 (2015): 76-84.

\bibitem{ref:izhikevich_improc} Izhikevich, Eugene M., Botond Szatmary, and Csaba Petre. "Invariant pulse latency coding systems and methods systems and methods." U.S. Patent 8,467,623, issued June 18, 2013.

\bibitem{ref:izhikevich_spike} Izhikevich, Eugene M. "Polychronization: computation with spikes." Neural computation 18, no. 2 (2006): 245-282.

\bibitem{ref:spike_querlioz} Querlioz, Damien, Olivier Bichler, and Christian Gamrat. "Simulation of a memristor-based spiking neural network immune to device variations." In Neural Networks (IJCNN), The 2011 International Joint Conference on, pp. 1775-1781. IEEE, 2011.

\bibitem{ref:essderc} Csaba, Gyorgy, Adam Papp, Wolfgang Porod, and Ramazan Yeniceri. "Non-boolean computing based on linear waves and oscillators." In 2015 45th European Solid State Device Research Conference (ESSDERC), pp. 101-104. IEEE, 2015.

\bibitem{ref:STDP_cruz_albrecht} Cruz-Albrecht, Jose M., Michael W. Yung, and Narayan Srinivasa. "Energy-efficient neuron, synapse and STDP integrated circuits." IEEE transactions on biomedical circuits and systems 6, no. 3 (2012): 246-256.


\bibitem{ref:neumann_patent} "Non-linear capacitance or inductance switching, amplifying, and memory organs." U.S. Patent 2,815,488, issued December 3, 1957.

\bibitem{ref:goto} Goto E. The parametron, a digital computing element which utilizes parametric oscillation. Proceedings of the IRE. 1959 Aug;47(8):1304-16.

\bibitem{ref:parametron1} Mahboob I, Yamaguchi H. Bit storage and bit flip operations in an electromechanical oscillator. Nature nanotechnology. 2008 May 1;3(5):275-9.


\bibitem{ref:musasino} Muroga S, Takashima K. The parametron digital computer Musasino-1. Electronic Computers, IRE Transactions on. 1959 Sep(3):308-16.

\bibitem{ref:nikonov_benchmarking} Dmitri E. Nikonov, Ian A. Young, Overview of beyond-CMOS devices and a uniform methodology for their benchmarking, Proc. IEEE 101 (12) (2013) 2498–2533.

\bibitem{ref:rowch} Roychowdhury J. : Boolean Computation Using Self-Sustaining Nonlinear Oscillators. Proceedings of the IEEE. 2015 Nov;103(11):1958-69.

\bibitem{ref:phlogon_book} Wang, Tianshi, and Jaijeet Roychowdhury. "PHLOGON: Phase-based logic using oscillatory nano-systems." In International Conference on Unconventional Computation and Natural Computation, pp. 353-366. Springer, Cham, 2014.



\bibitem{ref:nanomech3} Mahboob, I., and H. Yamaguchi. "Bit storage and bit flip operations in an electromechanical oscillator." Nature nanotechnology 3, no. 5 (2008): 275-279.

\bibitem{ref:kiehl_phaselogic} Fahmy, Hossam AH, and Richard A. Kiehl. "Complete logic family using tunneling-phase-logic devices." In Microelectronics, 1999. ICM'99. The Eleventh International Conference on, pp. 153-156. IEEE, 1999.

\bibitem{ref:montecarlo} Csaba, György, Trond Ytterdal, and Wolfgang Porod. "Neural network based on parametrically-pumped oscillators." In Electronics, Circuits and Systems (ICECS), 2016 IEEE International Conference on, pp. 45-48. IEEE, 2016.

\bibitem{ref:moore} see e.g. IEEE Spectrum: Special report: 50 years of Moore's law, The glorious history and inevitable decline of one of technology's greatest winning streaks http://spectrum.ieee.org/static/special-report-50-years-of-moores- law

\bibitem{ref:oscillator_book} Westra, Jan Roelof, Chris JM Verhoeven, and Arthur HM Van Roermund. Oscillators and Oscillator Systems. Kluwer, 2000. 

\bibitem{ref:lowpowerVCO2}  Farzeen, Suzana, Guoyan Ren, and Chunhong Chen. "An ultra-low power ring oscillator for passive UHF RFID transponders." In Circuits and Systems (MWSCAS), 2010 53rd IEEE International Midwest Symposium on, pp. 558-561. IEEE, 2010.

\bibitem{ref:Shukla2014IEDM} Shukla, Nikhil, Abhinav Parihar, Matthew Cotter, Michael Barth, Xueqing Li, Nandhini Chandramoorthy, Hanjong Paik et al. "Pairwise coupled hybrid vanadium dioxide-MOSFET (HVFET) oscillators for non-boolean associative computing." In Electron Devices Meeting (IEDM), 2014 IEEE International, pp. 28-7. IEEE, 2014.

\bibitem{ref:inductors_excellent} Gardner, Donald S., Gerhard Schrom, Fabrice Paillet, Brice Jamieson, Tanay Karnik, and Shekhar Borkar. "Review of on-chip inductor structures with magnetic films." IEEE Transactions on Magnetics 45, no. 10 (2009): 4760-4766.

\bibitem{ref:josephsonsync1} Wiesenfeld, Kurt, Pere Colet, and Steven H. Strogatz. "Synchronization transitions in a disordered Josephson series array." Physical review letters 76, no. 3 (1996): 404.

\bibitem{ref:josephsonsync2} Segall, Ken, Matthew LeGro, Steven Kaplan, Oleksiy Svitelskiy, Shreeya Khadka, Patrick Crotty, and Daniel Schult. "Synchronization dynamics on the picosecond time scale in coupled Josephson junction neurons." Physical Review E 95, no. 3 (2017): 032220.

\bibitem{ref:nanomech_1} Guerra, Diego N., Adi R. Bulsara, William L. Ditto, Sudeshna Sinha, K. Murali, and P. Mohanty. "A noise-assisted reprogrammable nanomechanical logic gate." Nano letters 10, no. 4 (2010): 1168-1171.

\bibitem{ref:grollier_spintorque} Locatelli, N., V. Cros, and J. Grollier. "Spin-torque building blocks." Nature Materials 13, no. 1 (2014).

\bibitem{ref:kaka_pufall} Kaka, Shehzaad, Matthew R. Pufall, William H. Rippard, Thomas J. Silva, Stephen E. Russek, and Jordan A. Katine. "Mutual phase-locking of microwave spin torque nano-oscillators." Nature 437, no. 7057 (2005): 389-392.

\bibitem{ref:chemical} Mazouz, Nadia, Katharina Krischer, Georg Flätgen, and Gerhard Ertl. "Synchronization and pattern formation in electrochemical oscillators: Model calculations." The Journal of Physical Chemistry B 101, no. 14 (1997): 2403-2410.

\bibitem{ref:krivorotov} Chen, Yu-Jin, Han Kyu Lee, Roman Verba, Jordan A. Katine, Igor Barsukov, Vasil Tiberkevich, John Q. Xiao, Andrei N. Slavin, and Ilya N. Krivorotov. "Parametric resonance of magnetization excited by electric field." Nano letters 17, no. 1 (2016): 572-577.



\bibitem{ref:akerman9sto} Awad, A. A., P. Drrenfeld, A. Houshang, M. Dvornik, E. Iacocca, R. K. Dumas, and Johan Akerman. "Long-range mutual synchronization of spin Hall nano-oscillators." Nat. Phys. (2016).

\bibitem{ref:nanomech_2} Mahboob, I., E. Flurin, K. Nishiguchi, A. Fujiwara, and H. Yamaguchi. "Interconnect-free parallel logic circuits in a single mechanical resonator." Nature communications 2 (2011): 198.

\bibitem{ref:nanomech_4} Weinstein, Dana, and Sunil A. Bhave. "The resonant body transistor." Nano letters 10, no. 4 (2010): 1234-1237.

\bibitem{ref:mechanical_new} Coulombe, Jean C., Mark CA York, and Julien Sylvestre. "Computing with networks of nonlinear mechanical oscillators." PloS one 12, no. 6 (2017): e0178663.

\bibitem{ref:highQcmos} Nguyen, CT-C., and Roger T. Howe. "An integrated CMOS micromechanical resonator high-Q oscillator." IEEE Journal of Solid-State Circuits 34, no. 4 (1999): 440-455.

\bibitem{ref:mems_efficiency} Oliver Paul: Micro transducer operation, in.: Korvink, J., and Paul, O. (2010). MEMS: A practical guide of design, analysis, and applications. Springer Science  Business Media. Chicago

\bibitem{ref:rbo_new} Bahr, Bichoy, Yanbo He, Zoran Krivokapic, Srinivasa Banna, and Dana Weinstein. "32GHz resonant-fin transistors in 14nm FinFET technology." In Solid-State Circuits Conference-(ISSCC), 2018 IEEE International, pp. 348-350. IEEE, 2018.

\bibitem{ref:supercond00} Mage, Jean-Claude, Bruno Marcilhac, Maurice Poulain, Yves Lemaitre, Julien Kermorvant, and Jean-Marc Lesage. "Low noise oscillator based on 2D superconducting resonator." In Frequency Control and the European Frequency and Time Forum (FCS), 2011 Joint Conference of the IEEE International, pp. 1-4. IEEE, 2011.

\bibitem{ref:supercond_oscillator} Zhao, Jie, Peng Zhao, Haifeng Yu, and Yang Yu. "External driving synchronization in a superconducting quantum interference device based oscillator." Japanese Journal of Applied Physics 55, no. 11 (2016): 110301.

\bibitem{ref:ROneuron} Sahoo, Bibhu Datta. "Ring oscillator based sub-1V leaky integrate-and-fire neuron circuit." In Circuits and Systems (ISCAS), 2017 IEEE International Symposium on, pp. 1-4. IEEE, 2017.

\bibitem{ref:ringoscillator_old} Feuer, M. D., R. H. Hendel, R. A. Kiehl, J. C. M. Hwang, V. G. Keramidas, C. L. Allyn, and R. Dingle. "High-speed low-voltage ring oscillators based on selectively doped heterojunction transistors." IEEE Electron Device Letters 4, no. 9 (1983): 306-307.

\bibitem{ref:corinto_memristor} Corinto, Fernando, Alon Ascoli, and Marco Gilli. "Nonlinear dynamics of memristor oscillators." IEEE Transactions on Circuits and Systems I: Regular Papers 58, no. 6 (2011): 1323-1336.

\bibitem{ref:sto_1}  Ruotolo, Antonio, V. Cros, B. Georges, A. Dussaux, J. Grollier, C. Deranlot, R. Guillemet, K. Bouzehouane, S. Fusil, and A. Fert. "Phase-locking of magnetic vortices mediated by antivortices." Nature nanotechnology 4, no. 8 (2009): 528-532.

\bibitem{ref:parametric_magnet} Urazhdin, Sergei, Vasil Tiberkevich, and Andrei Slavin. "Parametric excitation of a magnetic nanocontact by a microwave field." Physical review letters 105, no. 23 (2010): 237204.

\bibitem{ref:lowpowerVCO} Deen, M. Jamal, Mehdi H. Kazemeini, and Sasan Naseh. "Performance characteristics of an ultra-low power VCO." In Circuits and Systems, 2003. ISCAS'03. Proceedings of the 2003 International Symposium on, vol. 1, pp. I-I. IEEE, 2003.

\bibitem{ref:parihar} Parihar, Abhinav, Nikhil Shukla, Suman Datta, and Arijit Raychowdhury. "Exploiting synchronization properties of correlated electron devices in a non-boolean computing fabric for template matching." IEEE Journal on Emerging and Selected Topics in Circuits and Systems 4, no. 4 (2014): 450-459.


\bibitem{ref:RO_phasenoise} Abidi, Asad A. "Phase noise and jitter in CMOS ring oscillators." IEEE Journal of Solid-State Circuits 41, no. 8 (2006): 1803-1816.

\bibitem{ref:maffezoni_models} Maffezzoni, Paolo, Bichoy Bahr, Zheng Zhang, and Luca Daniel. "Oscillator array models for associative memory and pattern recognition." IEEE Transactions on Circuits and Systems I: Regular Papers 62, no. 6 (2015): 1591-1598.

\bibitem{ref:sumandatta} Shukla N, Parihar A, Freeman E, Paik H, Stone G, Narayanan V, Wen H, Cai Z, Gopalan V, Engel-Herbert R, Schlom DG. Synchronized charge oscillations in correlated electron systems. Scientific reports. 2014;4.

\bibitem{ref:nikonov01} Nikonov DE, Csaba G, Porod W, Shibata T, Voils D, Hammerstrom D, Young IA, Bourianoff GI. Coupled-Oscillator Associative Memory Array Operation for Pattern Recognition. IEEE Journal on Exploratory Solid-State Computational Devices and Circuits. 2015 Dec;1:85-93.

\bibitem{ref:hanggi} Hänggi, Peter. "Stochastic resonance in biology how noise can enhance detection of weak signals and help improve biological information processing." ChemPhysChem 3, no. 3 (2002): 285-290.


\bibitem{ref:grollier_stochastic} Mizrahi, Alice, Nicolas Locatelli, Romain Lebrun, Vincent Cros, Akio Fukushima, Hitoshi Kubota, Shinji Yuasa, Damien Querlioz, and Julie Grollier. "Controlling the phase locking of stochastic magnetic bits for ultra-low power computation." Scientific reports 6 (2016).


\bibitem{ref:csaba_pufall_ieee} Csaba, Gyorgy, Wolfgang Porod, Matt Pufall, and William Rippard. "Analog circuits based on the synchronization of field-line coupled spin-torque oscillators." In Nanotechnology (IEEE-NANO), 2015 IEEE 15th International Conference on, pp. 1343-1345. IEEE, 2015.

\bibitem{ref:electrical_sto} Lebrun, R., S. Tsunegi, P. Bortolotti, H. Kubota, A. S. Jenkins, M. Romera, K. Yakushiji et al. "Mutual synchronization of spin torque nano-oscillators through a long-range and tunable electrical coupling scheme." Nature Communications 8 (2017): ncomms15825.

\bibitem{ref:mems_acoustic} Xu, Yuanjie, and Joshua EY Lee. "Mechanically coupled SOI Lamé-mode resonator-arrays: Synchronized oscillations with high quality factors of 1 million." In European Frequency and Time Forum and International Frequency Control Symposium (EFTF/IFC), 2013 Joint, pp. 133-136. IEEE, 2013.




\bibitem{ref:mems_hoppensteadt} Hoppensteadt, Frank C., and Eugene M. Izhikevich. "Synchronization of MEMS resonators and mechanical neurocomputing." IEEE Transactions on Circuits and Systems I: Fundamental Theory and Applications 48, no. 2 (2001): 133-138.

\bibitem{ref:hoppensteadt_pll} Hoppensteadt, Frank C., and Eugene M. Izhikevich. "Pattern recognition via synchronization in phase-locked loop neural networks." IEEE Transactions on Neural Networks 11, no. 3 (2000): 734-738.





\bibitem{ref:pikovsky} Pikovsky, Arkady, Michael Rosenblum, and Jürgen Kurths. Synchronization: a universal concept in nonlinear sciences. Vol. 12. Cambridge university press, 2003.


\bibitem{ref:kuramoto} Acebrón, Juan A., Luis L. Bonilla, Conrad J. Pérez Vicente, Félix Ritort, and Renato Spigler. "The Kuramoto model: A simple paradigm for synchronization phenomena." Reviews of modern physics 77, no. 1 (2005): 137.


\bibitem{ref:kuramoto_big} Strogatz, Steven H. "From Kuramoto to Crawford: exploring the onset of synchronization in populations of coupled oscillators." Physica D: Nonlinear Phenomena 143, no. 1 (2000): 1-20.

\bibitem{ref:izhikevich_weakly} Hoppensteadt, Frank C., and Eugene M. Izhikevich. Weakly connected neural networks. Vol. 126. Springer Science and Business Media, 2012.

\bibitem{ref:levitan_phase_model} Fang, Yan, Victor V. Yashin, Donald M. Chiarulli, and Steven P. Levitan. "A simplified phase model for oscillator based computing." In VLSI (ISVLSI), 2015 IEEE Computer Society Annual Symposium on, pp. 231-236. IEEE, 2015.

\bibitem{ref:suppes} Vassilieva, Ekaterina, Guillaume Pinto, José Acacio De Barros, and Patrick Suppes. "Learning pattern recognition through quasi-synchronization of phase oscillators." IEEE Transactions on Neural Networks 22, no. 1 (2011): 84-95.

\bibitem{ref:querlioz} Vodenicarevic, Damir, Nicolas Locatelli, Flavio Abreu Araujo, Julie Grollier, and Damien Querlioz. "A Nanotechnology-Ready Computing Scheme based on a Weakly Coupled Oscillator Network." Scientific Reports 7 (2017).

\bibitem{ref:izhikevich} Hoppensteadt FC, Izhikevich EM. Oscillatory neurocomputers with dynamic connectivity. Physical Review Letters. 1999 Apr 5;82(14):2983.

\bibitem{ref:neural_reservoir} Yamane T., Katayama Y., Nakane R., Tanaka G., Nakano D. (2015) Wave-Based Reservoir Computing by Synchronization of Coupled Oscillators. In: Arik S., Huang T., Lai W., Liu Q. (eds) Neural Information Processing. Lecture Notes in Computer Science, vol 9491. Springer, Cham

\bibitem{ref:hopfield} Hopfield, John J. "Neural networks and physical systems with emergent collective computational abilities." Proceedings of the national academy of sciences 79, no. 8 (1982): 2554-2558.

\bibitem{ref:porod_hopfield} Michel, Anthony N., Jay A. Farrell, and Wolfgang Porod. "Qualitative analysis of neural networks." IEEE Transactions on Circuits and Systems 36, no. 2 (1989): 229-243.

\bibitem{ref:bogdan} Bogdan Popescu, Gyorgy Csaba, Dan Popescu, Amir Hossein Fallahpour, Paolo Lugli, Wolfgang Porod, Markus Becherer: Simulation of Coupled Spin Torque Oscillators for Pattern Recognition submmited to IEEE Trans Nanotech 2017

\bibitem{ref:baldi} Baldi P, Meir R. Computing with arrays of coupled oscillators: An application to preattentive texture discrimination. Neural Computation. 1990;2(4):458-71.


\bibitem{ref:furber_temple} Furber, Steve, and Steve Temple. "Neural systems engineering." Journal of the Royal Society interface 4, no. 13 (2007): 193-206.

\bibitem{ref:superimportant} Schuman, Catherine D., Thomas E. Potok, Robert M. Patton, J. Douglas Birdwell, Mark E. Dean, Garrett S. Rose, and James S. Plank. "A Survey of Neuromorphic Computing and Neural Networks in Hardware." arXiv preprint arXiv:1705.06963 (2017).

\bibitem{ref:largescalebrain} Hugo de Garis, Chen Shuo, Ben Goertzel, Lian Ruiting: A world survey of artificial brain projects, Part I: Large-scale brain simulations Neurocomputing 74 (2010) 3?29

\bibitem{ref:yuminaka} Yuminaka, Yasushi, Yoshisato Sasaki, Takafumi Aoki, and Tatsuo Higuchi. "Design of neural networks based on wave-parallel computing technique." In Cellular Neural Networks and Analog VLSI, pp. 91-103. Springer US, 1998.

\bibitem{ref:fm_cnn} Mondragon-Torres, Antonio, Raman Gonzalez-Carvajal, J. Pineda de Gyvez, and Edgar Sanchez-Sinencio. "Frequency-domain intrachip communication schemes for CNN." In Cellular Neural Networks and Their Applications Proceedings, 1998 Fifth IEEE International Workshop on, pp. 398-403. IEEE, 1998.

\bibitem{ref:fm_neural} Craven, M. P., K. M. Curtis, and B. R. Hayes-Gill. "Frequency division multiplexing in analogue neural network." Electronics Letters 27, no. 11 (1991): 918-920.

\bibitem{ref:fm_neural2} Craven, M. P., K. M. Curtis, and B. R. Hayes-Gill. "Consideration of multiplexing in neural network hardware." IEE Proceedings-Circuits, Devices and Systems 141, no. 3 (1994): 237-240.

\bibitem{ref:heger_krischer_new} Heger, Daniel, and Katharina Krischer. "Robust autoassociative memory with coupled networks of Kuramoto-type oscillators." Physical Review E 94, no. 2 (2016): 022309.


\bibitem{ref:horvath} Horvath, Andras, Gyorgy Csaba, and Wolfgang Porod. "Dynamic coupling of spin torque oscillators for associative memories." In 2014 14th International Workshop on Cellular Nanoscale Networks and their Applications (CNNA), pp. 1-2. IEEE, 2014.




\bibitem{ref:marculescu} Marculescu, Radu, Umit Y. Ogras, Li-Shiuan Peh, Natalie Enright Jerger, and Yatin Hoskote. "Outstanding research problems in NoC design: system, microarchitecture, and circuit perspectives." IEEE Transactions on Computer-Aided Design of Integrated Circuits and Systems 28, no. 1 (2009): 3-21.

\bibitem{ref:ann_book} Graupe, Daniel. Principles of artificial neural networks. Vol. 7. World Scientific, 2013.

\bibitem{ref:nn_book2} Anderson, Dave, and George McNeill. "Artificial neural networks technology." Kaman Sciences Corporation 258, no. 6 (1992): 1-83.

\bibitem{ref:goodfellow_deepleraning} Goodfellow, Ian, Yoshua Bengio, and Aaron Courville. Deep learning. MIT press, 2016.

\bibitem{ref:dengdongdeep} Yu, Dong, and Li Deng. "Deep learning and its applications to signal and information processing [exploratory dsp]." IEEE Signal Processing Magazine 28, no. 1 (2011): 145-154.

\bibitem{ref:upside_wired} Darpa Has Seen the Future of Computing ... And It's Analog https://www.wired.com/2012/08/upside/ also https://www.darpa.mil/program/unconventional-processing-of-signals-for- intelligent-data-exploitation

\bibitem{ref:tadashi} Bui, Trong Tu, and Tadashi Shibata. "Compact bell-shaped analog matching-cell module for digital-memory-based associative processors." Japanese Journal of Applied Physics 47, no. 4S (2008): 2788.

\bibitem{ref:kaushikroy} Yogendra, Karthik, Chamika Liyanagedera, Deliang Fan, Yong Shim, and Kaushik Roy. "Coupled Spin-Torque Nano-Oscillator-Based Computation: A Simulation Study." ACM Journal on Emerging Technologies in Computing Systems (JETC) 13, no. 4 (2017): 56.

\bibitem{ref:csabasto} Csaba G, Porod W. Computational study of spin-torque oscillator interactions for non-Boolean computing applications. IEEE Transactions on Magnetics. 2013 Jul;49(7):4447-51.

\bibitem{ref:pufall_new} M. Pufall, W.H. Rippard, E. Jue, G. Csaba and K. Roy:  Estimating Degree of Match with Arrays of Spin Torque Oscillators. (Invited)  presented a

\bibitem{ref:syncspecial} Follmann, Rosangela, Elbert EN Macau, Epaminondas Rosa, and José RC Piqueira. "Phase oscillatory network and visual pattern recognition." IEEE transactions on neural networks and learning systems 26, no. 7 (2015): 1539-1544.

\bibitem{ref:grollier_new2} Romera, Miguel, Philippe Talatchian, Sumito Tsunegi, Flavio Abreu Araujo, Vincent Cros, Paolo Bortolotti, Kay Yakushiji et al. "Training coupled spin-torque nano-oscillators to classify patterns in real-time." arXiv preprint arXiv:1711.02704 (2017).

\bibitem{ref:karp21} Karp, Richard M. "Reducibility among combinatorial problems." In Complexity of computer computations, pp. 85-103. springer US, 1972.



\bibitem{ref:turinglimit} Siegelmann, Hava T. "Computation beyond the Turing limit." In Neural Networks and Analog Computation, pp. 153-164. Birkhauser Boston, 1999.

\bibitem{ref:dwave} M. W. Johnson,	M. H. S. Amin,	S. Gildert,	T. Lanting,	F. Hamze,	N. Dickson,	R. Harris,	A. J. Berkley,	J. Johansson,	P. Bunyk,	E. M. Chapple,	C. Enderud,	J. P. Hilton,	K. Karimi,	E. Ladizinsky,	N. Ladizinsky,	T. Oh,	I. Perminov,	C. Rich,	M. C. Thom,	E. Tolkacheva,	C. J. S. Truncik,	S. Uchaikin,	J. Wang,	B. Wilson	et al. Quantum annealing with manufactured spins Nature 473, 194–198 (12 May 2011) doi:10.1038/nature10012


\bibitem{ref:aaronson} Aaronson, Scott. "Guest column: NP-complete problems and physical reality." ACM Sigact News 36, no. 1 (2005): 30-52.

\bibitem{ref:graphcoloring_earliest} Wu, Chai Wah. "Graph coloring via synchronization of coupled oscillators." IEEE Transactions on Circuits and Systems I: Fundamental Theory and Applications 45, no. 9 (1998): 974-978.

\bibitem{ref:wu_graphcoloring} Wu, Jianshe, Licheng Jiao, Rui Li, and Weisheng Chen. "Clustering dynamics of nonlinear oscillator network: Application to graph coloring problem." Physica D: Nonlinear Phenomena 240, no. 24 (2011): 1972-1978.

\bibitem{ref:memcomputing_1} Traversa, Fabio Lorenzo, Chiara Ramella, Fabrizio Bonani, and Massimiliano Di Ventra. "Memcomputing NP-complete problems in polynomial time using polynomial resources and collective states." Science advances 1, no. 6 (2015): e1500031.

\bibitem{ref:memcomputing_2} Pershin, Yuriy V., and Massimiliano Di Ventra. "Memcomputing: A computing paradigm to store and process information on the same physical platform." In Computational Electronics (IWCE), 2014 International Workshop on, pp. 1-2. IEEE, 2014.

\bibitem{ref:memcomputing_3} Di Ventra, Massimiliano, and Yuriy V. Pershin. "Memcomputing: a computing paradigm to store and process information on the same physical platform." arXiv preprint arXiv:1211.4487 (2012).

\bibitem{ref:memcomputing_4} Traversa, Fabio Lorenzo, and Massimiliano Di Ventra. "Universal memcomputing machines." IEEE transactions on neural networks and learning systems 26, no. 11 (2015): 2702-2715.

\bibitem{ref:memcomputing_sciam} Massimiliano Di Ventra and Yuriy V. Pershin  Just Add Memory Scientific American 312, 56 - 61 (2015) Published online: 20 January 2015

\bibitem{ref:memcomputing05} Pershin, Yuriy V., and Massimiliano Di Ventra. "Solving mazes with memristors: A massively parallel approach." Physical Review E 84, no. 4 (2011): 046703.

\bibitem{ref:memcomputing06} Traversa, Fabio L., Fabrizio Bonani, Yuriy V. Pershin, and Massimiliano Di Ventra. "Dynamic computing random access memory." Nanotechnology 25, no. 28 (2014): 285201.

\bibitem{ref:cnn_np_ercsey} Ercsey-Ravasz, Maria, Tamas Roska, and Zoltan Neda. "Cellular neural networks for NP-hard optimization." EURASIP Journal on Advances in Signal Processing 2009 (2009): 2.

\bibitem{ref:ercsey_sudoku} Ercsey-Ravasz, Maria, and Zoltan Toroczkai. "The chaos within Sudoku." Scientific reports 2 (2012): 725.

\bibitem{ref:ercsey_ksat} Maria Ercsey-Ravasz and Zoltan Toroczkai Optimization hardness as transient chaos in an analog approach to constraint satisfaction Nature Physics | Vol 7 | December 2011 |

\bibitem{ref:brocket_analog} Brockett R.W. (1992) Analog and digital computing. In: Bensoussan A., Verjus J.P. (eds) Future Tendencies in Computer Science, Control and Applied Mathematics. Lecture Notes in Computer Science, vol 653. Springer, Berlin, Heidelberg

\bibitem{ref:winograd} Winograd, Shmuel, and Jack D. Cowan. Reliable computation in the presence of noise. No. 22. Cambridge, Mass.: MIT Press, 1963.


\end{thebibliography}
\end{document}